\newcommand{\beq}{\begin{equation}}
\newcommand{\eeq}{\end{equation}}
\newcommand{\bea}{\begin{eqnarray}}
\newcommand{\eea}{\end{eqnarray}}

\newcommand{\gsim}{\lower.7ex\hbox{$\;\stackrel{\textstyle>}{\sim}\;$}}
\newcommand{\lsim}{\lower.7ex\hbox{$\;\stackrel{\textstyle<}{\sim}\;$}}

\newcommand{\mrm}{\mathrm}

\documentclass[aps,prev,twocolumn,preprintnumbers,floatfix,nofootinbib]{revtex4}
\usepackage{graphicx}
\usepackage{epstopdf}
\usepackage{mathrsfs}
\usepackage{amssymb}
\usepackage{verbatim}

\def\stacksymbols #1#2#3#4{\def\theguybelow{#2}
    \def\vp{\lower#3pt}
    \def\sp{\baselineskip0pt\lineskip#4pt}
    \mathrel{\mathpalette\intermediary#1}}

\def\intermediary#1#2{\vp\vbox{\sp
     \everycr={}\tabskip0pt
     \halign{$\mathsurround0pt#1\hfil##\hfil$\crcr#2\crcr
              \theguybelow\crcr}}}

\def\be{\begin{equation}}
\def\ee{\end{equation}}
\def\bea{\begin{eqnarray}}
\def\eea{\end{eqnarray}}

\def\sp{\;\;\;,\;\;\;}

\def\mrm{\mathrm}

\def\lsim{\raise0.3ex\hbox{$\;<$\kern-0.75em\raise-1.1ex\hbox{$\sim\;$}}}
\def\gsim{\raise0.3ex\hbox{$\;>$\kern-0.75em\raise-1.1ex\hbox{$\sim\;$}}}

\def\inbar{\,\vrule height1.5ex width.4pt depth0pt}

\def\IC{\relax\hbox{$\inbar\kern-.3em{\rm C}$}}
\def\IQ{\relax\hbox{$\inbar\kern-.3em{\rm Q}$}}
\def\IR{\relax{\rm I\kern-.18em R}}
 \font\cmss=cmss10 \font\cmsss=cmss10 at 7pt
\def\IZ{\relax\ifmmode\mathchoice
 {\hbox{\cmss Z\kern-.4em Z}}{\hbox{\cmss Z\kern-.4em Z}}
 {\lower.9pt\hbox{\cmsss Z\kern-.4em Z}}
 {\lower1.2pt\hbox{\cmsss Z\kern-.4em Z}}\else{\cmss Z\kern-.4em Z}\fi}

\def\comment#1{}
\def\to{\rightarrow}

\def\u1x{U(1)_X}
\newcommand{\nc}{\newcommand}
\nc{\LL}{L}
\nc{\vv}{\tilde{v}}
\nc{\ccdot}{\!\cdot\!}
\nc{\gsm}{G_{SM}}
\nc{\vfive}{\mathbf{5}\oplus\mathbf{\overline{5}}}
\nc{\vten}{\mathbf{10}\oplus\mathbf{\overline{10}}}
\nc{\zhol}{Z^{\rm hol}}
\nc{\xfb}{\,{\rm fb}}

\begin{document}

%
%

\preprint{LPT--Orsay 12-42}

\title{Direct detection of Higgs--portal dark matter at the LHC}

\author{Abdelhak Djouadi$^{a,b}$}
\email{abdelhak.djouadi@th.u-psud.fr}
\author{Adam Falkowski$^{a}$}
\email{adam.falkowski@th.u-psud.fr}
\author{Yann Mambrini$^{a}$}
\email{yann.mambrini@th.u-psud.fr}
\author{J\'er\'emie Quevillon$^{a}$}
\email{jeremie.quevillon@th.u-psud.fr}

\vspace{0.1cm}
\affiliation{
\mbox{$^a$ Laboratoire de Physique Th\'eorique, Universit\'e Paris-Sud, F-91405
Orsay, France.}
\\ 
${}^b$ CERN, CH--1211, Geneva 23, Switzerland.
}

\begin{abstract} 

We consider the process in which a Higgs particle is produced in association with jets and show that monojet searches at the LHC already provide interesting constraints on the invisible decays of a 125 GeV Higgs boson. Using the existing monojet searches performed by  CMS and ATLAS,  we show the 95\% confidence level limit on the invisible Higgs decay rate is of the order of the total Higgs production rate in the Standard Model. 
This limit could be significantly improved when more data at higher center of mass energies are collected,  provided systematic errors on the Standard Model contribution to the monojet background can be reduced. 
We also compare these direct constraints on the invisible rate with indirect ones based on measuring the Higgs rates in visible channels. 
In the context of Higgs portal models of dark matter, we then discuss  how the LHC limits on the invisible Higgs branching fraction impose strong constraints on the dark matter  scattering cross section  on nucleons probed in direct detection experiments.

  \end{abstract}

\maketitle


\section*{Introduction}

The existence of a boson with a mass around $M_H=125$ GeV is now firmly established \cite{higgs}. 
The observed properties of the new particle are consistent with those of the Standard Model (SM) Higgs boson \cite{higgsfits}. 
Nevertheless, it is conceivable that the Higgs particle may have other decay channels that are not predicted by the SM. 
Determining or constraining non-standard Higgs boson decays will provide a vital input to model building beyond the SM.     

A very interesting possibility that is often discussed is a Higgs boson decaying
into stable particles that do not interact with the detector.
Common examples where Higgs particles can
have invisible decay modes include decays into the lightest supersymmetric particle \cite{Reviews}  or decays into heavy
neutrinos in the SM extended by a fourth  generation of fermions \cite{SM4}.  In a
wider context, the Higgs boson could be coupled to the particle that constitutes
all or part of the dark matter in the universe.  
In these so-called {\em Higgs portal}  models \cite{portal} the Higgs boson is the key mediator in the
process of dark matter annihilation and scattering, providing an intimate link
between Higgs hunting  in collider experiments and the direct search for dark
matter particles  in their elastic scattering on nucleons.  In fact,  the
present LHC Higgs search results, combined with the constraints on
the direct detection cross section from the XENON  experi\-ment
\cite{Aprile:2011ts},  severely constrain the Higgs couplings to dark matter
particles  and have strong consequences on invisible Higgs decay modes for
scalar,   fermionic  or vectorial  dark matter candidates
\cite{Djouadi:2011aa}. 

At the LHC, the main channel for producing a relatively light SM--like Higgs
boson is  the gluon--gluon fusion (ggF) mechanism. At leading order (LO), the process
proceeds through a heavy top quark loop, leading to a single Higgs boson in the
final state, $gg\to H$ \cite{ggH-LO}. A next-to-leading order  (NLO) in perturbative
QCD, an additional jet can be emitted by the initial  gluons or the internal
heavy quarks, leading to $gg\to Hg$ final states \cite{ggH-NLO} (additional
contributions are also provided by the $gq \to Hq$ process). As the QCD 
corrections turn out to be quite large,  the rate for $H\!+\!1\;$jet is not much
smaller than  the rate for $H\!+\!0\;$jet. The next-to-next-to-leading order (NNLO) QCD
corrections \cite{ggH-NNLO,deFlorian:2012mx}, besides significantly increasing
the $H\!+\!0$    and $H\!+\!1\;$jet rates, lead to $H\!+\!2\;$jet events. The
latter event topology  also occurs at LO in two other Higgs
production mechanisms: vector boson fusion (VBF) $qq\to Hqq$  and
Higgs--strahlung (VH) $q\bar q \to HW/HZ \to Hq\bar q$ which have rather
distinct kinematical features compared to the gluon fusion process; for a
review, see Ref.~\cite{Reviews}

Hence, if the Higgs boson is coupled to invisible particles, it may recoil against hard QCD radiation, leading to monojet events at the LHC. 
The potential of monojets searches to constrain the invisible decay width of a light Higgs boson has been pointed out before \cite{Bai:2011wz}. 
In this paper we update and extend these analyses.  We place constraints on the Higgs invisible rate defined as 
\begin{eqnarray}
\label{e.rinv}
R_{\rm inv}^{\rm pp} &= & {\sigma (p p  \to H) \times {\rm BR}  (H  \to {\rm inv.})\over \sigma (p p  \to H)_{SM} }.  
\end{eqnarray}
We will argue that the existing monojet searches at the LHC \cite{atlas_mono,cms_mono} yield the constrain $R_{\rm inv}^{\rm pp} \lesssim 1$.
The constraint is much better than expected.  
Indeed, early studies \cite{Hinvisible},  focusing mainly on the VBF production channel,  concluded that observation of invisible Higgs decays was only possible at the highest LHC  energy, $\sqrt s= 14$ TeV, and with more than 10 fb$^{-1}$ data.
Bounds on invisible Higgs  based on  the 1~fb$^{-1}$ monojet search in ATLAS \cite{Aad:2011xw} were studied in Ref. \cite{Bai:2011wz}, 
where a  weaker limit of $R_{\rm inv}^{\rm pp} \lsim 4$ was obtained for $M_h \sim 125$ GeV.  

One one hand, the constraint at the level $R_{\rm inv}^{\rm pp} \sim 1$ means that the monojet searches cannot yet significantly constrain the invisible Higgs branching fraction if the production rate of the 125 GeV Higgs boson is close to the SM one. 
In fact,  in that case much stronger constraints follow from global analyses of the {\em visible} Higgs decay channels, which disfavor ${\rm BR} (H \to {\rm inv.}) >  0.2$ at 95\% confidence level (CL) \cite{higgsfits}.    
However, in models beyond the SM,  the Higgs production rate may well be enhanced, and in that case the monojet constraints discussed here may become relevant. 
In this sense, our results are complementary to the indirect constraints on the invisible branching fraction obtained by measuring visible Higgs decays. 

In the next step, we discuss the connection between the Higgs invisible branching fraction and the direct dark matter detection cross section.  
 We work in the context of Higgs portal models and consider the cases of scalar, fermionic and vectorial dark matter
particles (that we generically denote by $\chi$) coupled to the Higgs boson.  To keep our discussion more general, the
Higgs--$\chi \chi$  couplings are not fixed by the requirement of obtaining the
correct relic density from thermal history\footnote{Instead, we assume that one
of the multiple possible  processes (e.g. co-annihilation, non-thermal production,
$s$--channel poles of particles from another sector)  could arrange that the
dark matter relic abundance is consistent with cosmological observations.}.   In
each case, the LHC constraint  ${\rm BR} (H \to {\rm inv.})$ can be translated into a constraint
on the Higgs boson couplings to the dark matter particles.  We will show that
these constraints are competitive with those derived from the XENON bounds on
the dark matter scattering cross section on nucleons\footnote{We note that the
process $gg\!\to\! H\!\to \!\chi \chi$  for dark matter $\chi$ production  at
the LHC is an important component  of the (crossed) process for dark matter
scattering on nucleons, $g \chi\! \to\! g \chi$ \cite{Manuel}.}.  We discuss how future
results from invisible Higgs searches at the LHC and from direct detection
experiments  will be complementary in exploring the parameter space of  Higgs
portal models.

The rest of this letter is organized as follows. 
In the next section, we  present our analysis of invisible Higgs production at the LHC.  
We estimate the sensitivity to the invisible Higgs rate of the CMS monojet search using 4.7 fb$^{-1}$ of data at $\sqrt s=7$ TeV  \cite{cms_mono}. 
We also study the constraints from  the recent  ATLAS monojet search using 10~fb$^{-1}$ of data at $\sqrt s=8$ TeV  \cite{atlas_mono}. 
In the following section we  discuss the interplay of the monojet constraints on the invisible Higgs decays and the indirect constraints from the global analysis of the LHC Higgs data.
We show that a portion of the theory space  with a large Higgs invisible branching fraction  favored by global fits is excluded by the monojet constraints. 
We then move on to discuss the implications for Higgs portal dark matter models and the complementarity between dark matter direct detection  at the LHC and in XENON. In the last section we present short conclusions.

\vspace*{-3mm}
\section*{Monojet constraints on the invisible width}
\vspace*{-3mm}

In this section we estimate the sensitivity of current monojet searches at the LHC to  a Higgs particles that decays invisibly.  
We rely on the searches  for monojets  performed by CMS using  4.7 fb$^{-1}$ of data at 7 TeV  center of mass energy \cite{cms_mono}. 
The basic selection requirements used by CMS  are as follows:\vspace*{-4mm}   

\begin{itemize} 
\item at least 1 jet with $p_T^j > 110$ GeV and $|\eta^j| < 2.4$;\vspace*{-3mm} 

\item at most 2 jets with $p_T^j > 30$ GeV;\vspace*{-3mm}   

\item no isolated leptons;\vspace*{-3mm} 

\end{itemize}

A second jet with $p_T^j$ above 30 GeV  is allowed provided it is not
back-to-back with the leading one, $\Delta \phi (j_1,j_2) < 2.5$.   Incidentally,
this is advantageous from the point of view of invisible Higgs searches, as
Higgs production at the LHC is often accompanied by more than one jet; vetoing
the second jet would reduce the signal acceptance by a factor of $\sim 2$.  The 
CMS collaboration quotes the observed event yields and expected  SM background for 4 different cuts on the missing
transverse momentum: $p_T^{\rm miss} > 250,300,350,400$~GeV.  
These events are largely dominated by the SM backgrounds, namely  $Z$+jets, where the $Z$ boson decays invisibly, and $W$+jets, where the $W$ boson decays leptonically and the charged lepton is not reconstructed. 
In particular, with 4.7 fb${}^{-1}$ data, the CMS collaboration estimates the background to be $7842 \pm 367$ events for $p_T^{\rm miss} > 250$ GeV. 

\begin{figure}[!h]
    \begin{center}
    \hspace{-1.cm}
   \includegraphics[width=2.4in]{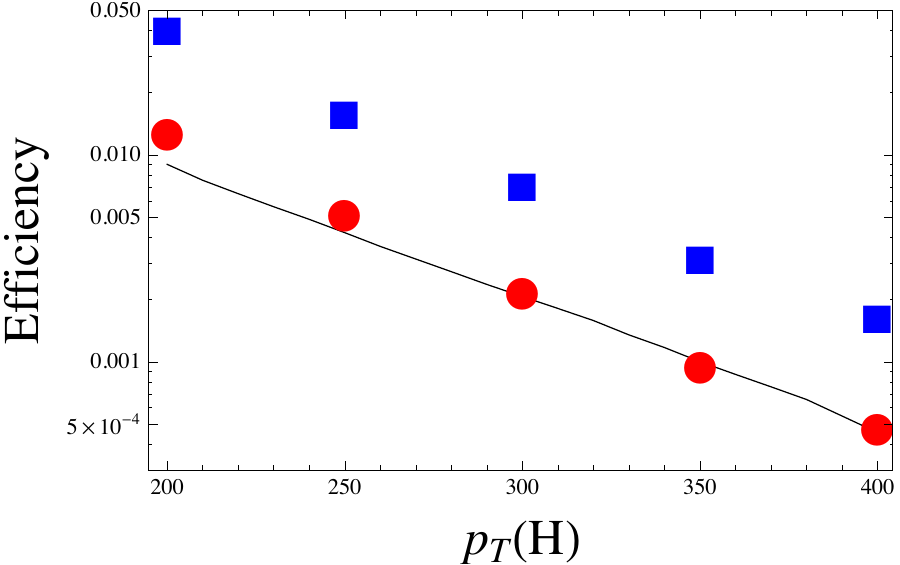}
\vspace*{-2mm}
          \caption{\footnotesize The fraction of events with Higgs transverse momentum 
	  above a given threshold for the ggF (red circles) and VBF (blue squares) 
	  production modes. The distributions were obtained at NLO using the
	  program  {\tt POWHEG}~\cite{Alioli:2008tz}. In the case of ggF, the simulations 
	  included the finite quark mass  effects  \cite{Bagnaschi:2011tu}, and we find good agreement with the NNLO distribution obtained using the  program {\tt HRes}~\cite{deFlorian:2012mx} (black line). }
\label{Fig:pth}
\end{center}
\vspace*{-3mm}
\end{figure}

A Higgs boson produced with a significant transverse momentum and decaying to invi\-si\-ble particles may also contribute to the final state targeted by monojet searches.   
In Fig.\ref{Fig:pth}, we show the fraction of  Higgs events  produced at the parton level in the ggF and VBF processes with $p_T^{\rm Higgs}$ above a given threshold, assuming   $M_H = 125$ GeV. 
One observes  that about 0.5\% of ggF  events are produced with $p_T^{\rm Higgs} > 250$ GeV, while for  the VBF production processes that fraction is larger by a factor of $\sim 3$. In 4.7 fb$^{-1}$ data at $\sqrt{s} = 7$ TeV this corresponds to about 500 events, assuming the SM production cross sections. 
This suggests that if an invisible Higgs boson is  produced with rates that are comparable or larger than that of the SM Higgs boson, the monojet searches may already provide meaningful constraints. 
 
In order to estimate the sensitivity of the CMS monojet search to the invisible Higgs signal, we generated the $pp \! \to \!H \!+ \!{\rm jets} \to {\rm
invisible} \!+ \!{\rm jets}$ process.  We used the program  {\tt POWHEG}~\cite{Alioli:2008tz,Bagnaschi:2011tu} for the ggF and VBF channels at the parton level, and  {\tt Madgraph 5} \cite{Alwall:2011uj} for the VH channels.
Showering and hadronization was performed using {\tt Pythia 6}~\cite{Sjostrand:2006za}  and {\tt Delphes 1.9}~\cite{Ovyn:2009tx}  was employed
to simulate the CMS detector response. We imposed the analysis cuts listed above on the simulated events so as to find the signal efficiency.   
As a cross-check, we passed $(Z \to \nu \nu)$ + jets background events through the same simulation chain, 
obtaining efficiencies consistent within 15\% with the data--driven estimates of that background provided by CMS. 
 
The signal event yield  depends on the cross section in each Higgs production channel and  on the
Higgs branching fraction into invisible final states.  Thus, strictly speaking, the quantities that
are being constrained  by the CMS search are\footnote{Assuming custodial symmetry,  $R_{\rm inv}^{\rm VH} =  R_{\rm inv}^{\rm VBF} \equiv R_{\rm inv}^{\rm V}$.}  $R_{\rm inv}^{\rm gg}$ and $R_{\rm inv}^{\rm V}$ defined as 
\begin{eqnarray}
\label{e.rinv2}
R_{\rm inv}^{\rm gg}  &= & {\sigma (g g \to  H)  \times  {\rm BR}
(H \to {\rm inv})\over \sigma (g g \to H)_{SM} }, \\
R_{\rm inv}^{\rm V} &= & {\left [ \sigma (q q  \to H q q) + \sigma (q \bar q  \to V H) \right ]  \times {\rm BR} 
(H  \to {\rm inv})\over \sigma (q q \to H q q)_{SM} + \sigma (q \bar q  \to V H)_{SM}   }  \nonumber
\end{eqnarray} 
Currently available data do not allow us to independently constrain $R_{\rm
inv}^{\rm gg}$ and  $R_{\rm inv}^{\rm V}$.  Thus, for the sake of setting
limits, we assume that the proportions of ggF, VBF and VH rates are the same as
in the SM, and we take the inclusive cross sections to be  $\sigma(gg \to
H)_{\rm SM} = 15.3$~pb,  $\sigma(q q\to H q q)_{\rm SM} = 1.2$~pb and  $\sigma(q
\bar q \to H V)_{\rm SM} = 0.9$~pb \cite{LHCXS}.   With this assumption, after
the analysis cuts the signal receives about $30\%$ contribution from the VBF
and VH production modes,  and the rest from ggF;  thus CMS  constrains the
combination $R_{\rm inv}^{\rm pp} \approx \frac23  R_{\rm inv}^{\rm gg} + \frac13 R_{\rm inv}^{\rm V}$. 

Our results are presented in Table~\ref{Tab:nh}.  We display the predicted
event yields $N_{\rm inv}^{\rm gg}$, $N_{\rm inv}^{\rm V}$ in, respectively,  the ggF and VBF+VH channels for the four CMS $p_T^{\rm miss}$ cuts.\footnote{
Note that we did not consider the theoretical  uncertainties on the cross sections \cite{LHCXS} and the efficiencies of the
$p_T$ cuts  which, although significant, are currently smaller than the experimental  ones.}  
For convenience, we also reproduce the expected  $ \Delta N_{95 \%}^{i,\rm exp}$ and observed $ \Delta N_{95 \%}^{\rm obs}$  95\% CL limits on the number of extra non-SM events quoted by CMS in Ref.~\cite{cms_mono} for each cut.   
Comparing  $N_{\rm inv}^{\rm gg}+N_{\rm inv}^{\rm V}$ with $ \Delta N_{95 \%}$  it is straightforward to obtain 95\% CL expected and observed limits on  $R_{\rm inv}^{\rm pp}$ corresponding to each cut reported in Table~\ref{Tab:nh}. 
We find the best expected limit $R_{\rm inv}^{\rm pp} \leq 2.1$ for the $p_T^{\rm miss} \ge 250$~GeV cut.   
The observed limit is better than the expected one  thanks to an ${\cal O}(1 \sigma)$ downward fluctuation of the SM background, and we  find $R_{\rm inv}^{\rm pp} \leq 1.6$ at 95\% CL for that cut.  
A stronger limit on $R_{\rm inv}^{\rm pp}$  can be derived by binning the number of events given in Table~\ref{Tab:nh} into exclusive bins, and then combining exclusion limits from all  four $p_T^{\rm miss}$ bins. 
Assuming Gaussian errors, one can recast the limits on the number of non-SM events  as  
$\Delta N^i = \Delta N_{0}^i \pm \Delta N_{1 \sigma}^i$, with $\Delta N_{0}^i = \Delta N_{95 \%}^{i,\rm obs} - \Delta N_{95 \%}^{i,\rm exp}$,   $\Delta N_{1 \sigma}^i = \Delta N_{95 \%}^{i,\rm exp}/1.96$, where $i=1\dots 4$ indexes the  $p_T^{\rm miss}$ bins.   
Invisible Higgs decays would produce an excess of events $\delta N^i(R_{\rm inv}^{\rm pp})$ in all the bins. 
Assuming in addition  small correlations between the errors in various bins, we can thus construct a global $\chi^2$ function, 
$\chi^2 = \sum_i [ \Delta N_{0}^i - \delta N^i(R_{\rm inv}^{\rm pp})]^2/[\Delta N_{1 \sigma}^i]^2$ so as to constrain  $R_{\rm inv}^{\rm pp}$. 
Using this procedure we obtain 
\beq
R_{\rm inv}^{\rm pp} \leq 1.10  \quad {\rm at}\  95\% \ {\rm CL}.   
\eeq  
Following  the same  procedure, we can  also  constrain separately $R_{\rm inv}^{\rm gg}$ and  $R_{\rm inv}^{\rm V}$, assuming only the ggF or only the VBF+VH Higgs production mode is present.   
We find $R_{\rm inv}^{\rm gg} \leq 2.0$  (when VBF and VH are absent) or  $R_{\rm inv}^{\rm V} \leq 4.0$ (when ggF is absent) at 95\% CL.  

\begin{table}
\renewcommand{\arraystretch}{1.1}
\begin{tabular}{c|c|c|c|c|c|c}
$p_T^{\rm miss}$[GeV]  &$~~N_{\rm inv}^{\rm gg}$    & $N_{\rm inv}^{\rm V}$ & 
$\Delta N_{95 \%}^{\rm exp}$ & $\Delta N_{95 \%}^{\rm obs}$ & exp. $R_{\rm inv}^{\rm pp} $ & obs. $ R_{\rm inv}^{\rm pp}$ \\  \hline
250 & $ 250 $ &$  110 $  &  $779$ & $600$ &  2.1 & 1.6 \\  \hline
300 & $ 110 $ &$  50  $  & $325$ & $368$ & 2.1& 2.3 \\  \hline 
350 & $ 46  $ &$  25  $  &  $200$ & $158$ & 2.8 & 2.2  \\ \hline
400 & $ 22  $ &$  13  $  & $118$ & $95$ & 3.4 & 2.7 \\   \hline
\end{tabular}
\caption{\footnotesize Limits on the  on the invisible Higgs rate $R_{\rm inv}^{\rm pp}$. 
The event yields are given for each reported  $p_T^{\rm miss}$  cut  of the CMS monojet search, separately for the  ggF and VBF+VH production modes, assuming the SM Higgs production cross sections in these channels and ${\rm BR}  (H  \to {\rm inv}) = 100\%$.  
We also give the expected and observed 95\% CL limits on the number of non-SM events reported by CMS \cite{cms_mono}, which allow us to derive 95\%CL expected and  observed limits on  $R_{\rm inv}^{\rm pp}$. 
 }
\label{Tab:nh}
\vspace*{-3mm}
\end{table}

We also study the impact of the ATLAS monojet search  \cite{atlas_mono} with 10~fb$^{-1}$  at $\sqrt{s} = 8$~TeV. 
ATLAS defines 4 search categories: SR1, SR2, SR3, SR4 with similar cuts on the visible jets as discussed above for the CMS case,   
and with the missing energy cut $p_T^{\rm miss} >120, 220,350,500$~GeV, respectively. 
In Table~\ref{Tab:atlas}  we give the 95\% CL limits on the invisible rate deduced from the number  monojet events reported by ATLAS for each of these categories.    
We find the best expected limit $R_{\rm inv}^{\rm pp} \leq 1.7$ using the $p_T^{\rm miss} \ge 220$~GeV cut, while the best observed limit is $R_{\rm inv}^{\rm pp} \leq 1.4$ using the $p_T^{\rm miss} \ge 500$~GeV.  Unlike in the CMS case, combining ATLAS exclusion limits from different  $p_T^{\rm miss}$ bins does not improve the limit of $R_{\rm inv}^{\rm pp}$. 
\begin{table}
\renewcommand{\arraystretch}{1.1}
\begin{tabular}{c|cc|c|c|c}
$p_T^{\rm miss}$[GeV]  &$~~N_{\rm inv}^{\rm gg}$    & $N_{\rm inv}^{\rm V}$ & 
$\Delta N_{\rm Bkg}$ & exp. $R_{\rm inv}^{\rm pp} $ & obs. $R_{\rm inv}^{\rm pp}$ \\  \hline
120 & $ 5694$ &$  1543 $  & 12820 & 3.5 & 4.4 \\ \hline
220 & $ 904 $ &$  286$  &1030 & 1.7 & 1.6\\  \hline
350 & $110 $ &$  45 $  & 171 & 2.2& 3.3 \\  \hline 
500 & $ 15  $ &$  9  $  &  73 & 6.0 & 1.4  \\ \hline
\end{tabular}
\caption{\footnotesize Predicted event yields $N_{\rm inv}$ (assuming ${\rm BR}  (H  \to {\rm inv}) = 100\%$),  the $1 \sigma$ background uncertainty $\Delta N_{\rm Bkg}$,  and the expected and 
observed 95\% CL limits on the invisible Higgs rate $R_{\rm inv}^{\rm pp}$  for each reported missing energy cut in the 8 TeV 10 fb~$^{-1}$ ATLAS monojet search \cite{atlas_mono}. 
The event yields are given separately for the  ggF and VBF+VH production modes, assuming the SM Higgs production cross sections in these channels. 
 }
\label{Tab:atlas}
\vspace*{-3mm}
\end{table}

\vspace*{-3mm}
\section*{Monojet vs. indirect constraints on invisible decays}
\vspace*{-3mm}

\begin{figure}[!h]
    \begin{center}
    \hspace{-1.cm}
   \includegraphics[width=2.4in]{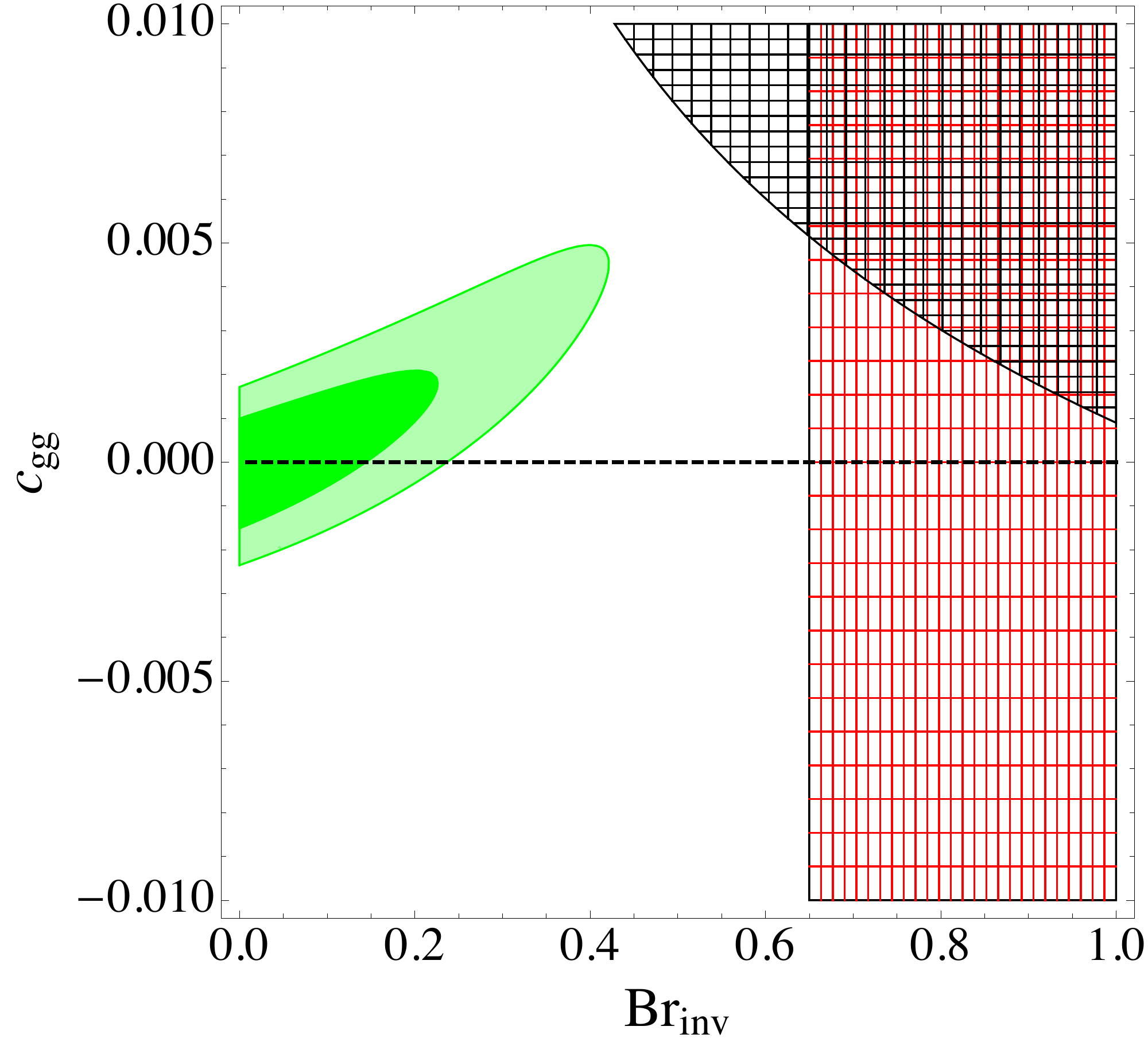}
\vspace*{-2mm}
          \caption{\footnotesize $68\%$ CL (light green) and $95\%$ CL (dark green) best fit regions to the combined LHC Higgs data.  
          The black meshed region is excluded by the monojet constraints derived in this paper,  while the red meshed region is  excluded by  the recent ATLAS $Z+(H \to {\rm MET})$ search \cite{ATLAS_Inv}.}
\label{Fig:fit}
\end{center}
\vspace*{-3mm}
\end{figure}

In this section we discuss the interplay between  the  monojet constraints on the invisible Higgs decays and the indirect constraints from the global analysis of the LHC Higgs data \cite{latesthiggs}. 
Assuming the Higgs is  produced with the SM cross section, the monojet constraints on the invisible branching fraction are not yet relevant.
However, in models beyond the SM the Higgs production rate can be significantly enhanced, especially in the gluon fusion channel. 
One well known example is the case of  the SM extended by the 4th generation of chiral fermions where the $g g \to H$ cross section is enhanced by an order of magnitude. In that class of models a large invisible width may easily arise due to Higgs decays to the 4th generation neutrinos, in which case the monojet constraints discussed here become very important.    
More generally,  the ggF rate can be enhanced whenever there exist additional colored scalars or fermions whose mass originates (entirely or in part) from electroweak symmetry breaking.  
In a model-independent way, we can describe their effect on the ggF rate via the effective Higgs coupling to gluons: 
\beq
\Delta {\cal L} = {c_{gg} \over 4} H G_{\mu \nu}^a G^{\mu \nu,a},
\eeq 
where $c_{gg}$ can take arbitrary real values depending on the number of additional colored species, their masses, their spins, and their couplings to the Higgs.
Furthermore, given the small Higgs width in the SM, $\Gamma_{H,SM} \sim 10^{-5} m_H$,  a significant invisible width $\Gamma_{H,\rm inv} \sim \Gamma_{H,SM}$ may easily arise even from small couplings of the Higgs to new physics, for example to massive neutrinos or to dark matter in Higgs portal models. 
We parametrize these possible couplings simply via the invisible branching fraction ${\rm Br}_{\rm inv}$, which is allowed to take  any value between 0 and 1. 
In Fig~\ref{Fig:fit} we plot the best fit region to the LHC Higgs data in the ${\rm Br}_{\rm inv}$-$c_{gg}$ parameter space. 
For the SM value $c_{gg}=0$ an invisible branching fraction larger than $\sim 20\%$ is disfavored at $95\%$ CL. 
 When $c_{gg} > 0$, the global fit admits a larger invisible branching fraction, even up to  ${\rm Br}_{\rm inv} \sim 50\%$.
 Nevertheless, the monojet constraints on the Higgs invisible width  derived in this paper are weaker then the indirect constraints from the global fits, when the latest Higgs data are taken into account. 

\vspace*{-3mm}
\section*{Invisible branching fraction and direct detection}
\vspace*{-3mm}

\begin{figure}[!h]
    \begin{center}
    \hspace{-1.cm}
   \includegraphics[width=3.in]{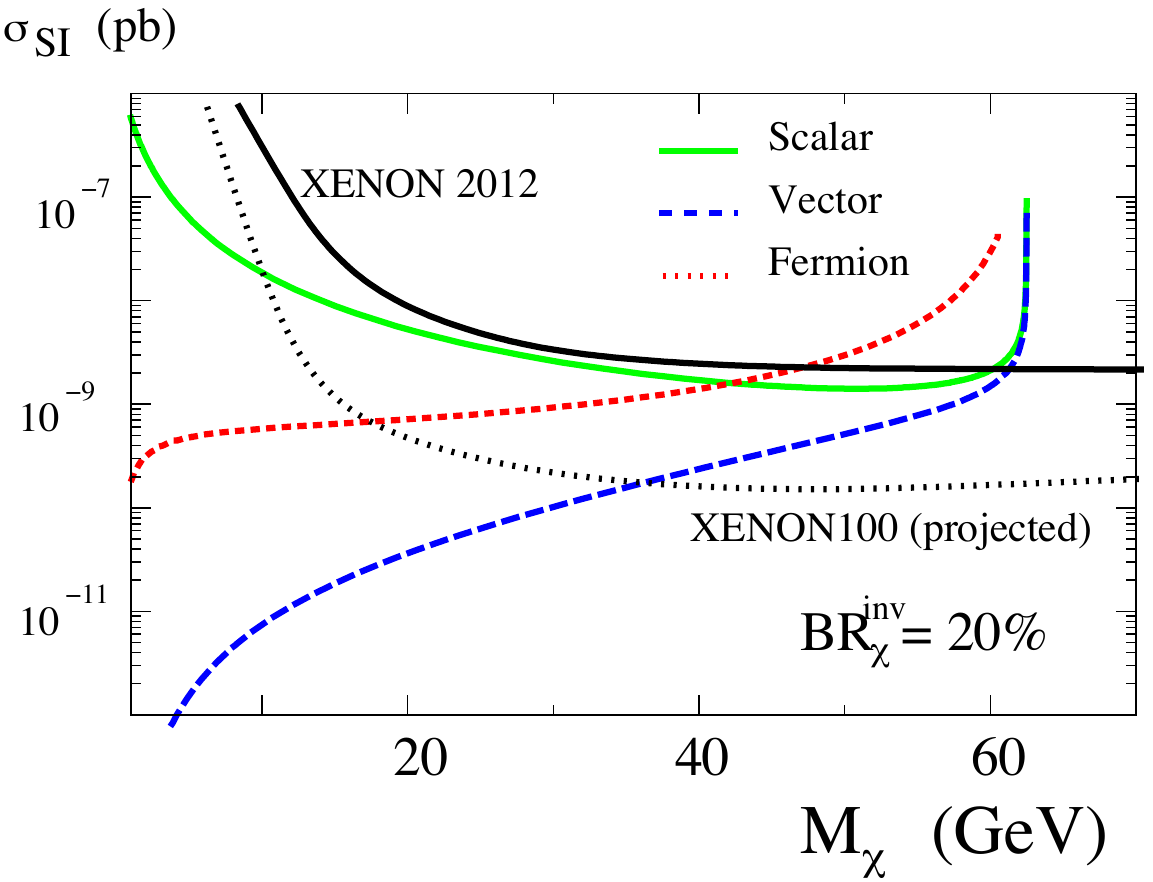}
\vspace*{-2mm}
          \caption{{\footnotesize
Bounds on the spin-independent direct detection cross section $\sigma^{\rm
SI}_{\chi p}$ in Higgs portal models derived for $M_H=125$ GeV and the 
invisible branching fraction of 20 \% (colored lines). The curves take into account the full $M_\chi$ dependence,  without using the approximation in Eq.~\ref{Eq:brmax}.  
For comparison, we plot the  current and future direct bounds from the 
XENON experiment (black lines). 
}}
\label{Fig:sigma40}
\end{center}
\vspace*{-9mm}
\end{figure}

If the invisible particle into which the Higgs boson decays is a constituent of dark matter in the universe, the Higgs coupling to dark matter can be probed
not  only at the LHC but also  in direct detection experiments.  In this
section, we  discuss the complementarity of these two direct detection methods. 
We consider generic Higgs-portal  scenarios in which the dark matter particle 
is a real scalar, a real vector, or a Majorana fermion,
$\chi=S,V,f$~\cite{Kanemura:2010sh,Djouadi:2011aa}.  The relevant  terms in the
effective Lagrangian in each of these cases are 
\begin{eqnarray} \!&&\Delta {\cal L}_S = -{1\over 2} m_S^2 S^2 - {1\over 4}
\lambda_S S^4 -  {1\over 4} \lambda_{hSS}  H^\dagger H  S^2 \;, \nonumber \\
\!&&\Delta {\cal L}_V = {1\over 2} m_V^2 V_\mu V^\mu\! +\! {1\over 4} \lambda_{V} 
(V_\mu V^\mu)^2\! +\! {1\over 4} \lambda_{hVV}  H^\dagger H V_\mu V^\mu ,
\nonumber \\ \! &&\Delta {\cal L}_f = - {1\over 2} m_f  f  f -  {1\over 4} {\lambda_{hff}\over \Lambda} H^\dagger H  f f + {\rm h.c.}  \;.   
\end{eqnarray}
The partial Higgs decay width into dark matter $\Gamma (H \to \chi\chi)$ and the
spin--independent  $\chi$--proton elastic cross section  $\sigma^{\rm SI}_{\chi p}$ can
be easily calculated in terms of the parameters of the Lagrangian, and we refer
to Ref.~\cite{Djouadi:2011aa}  for complete expressions.  For the present
purpose, it is important that both  $\Gamma (H \to \chi\chi)$ and   $\sigma^{\rm SI}_{\chi p}$ are proportional to $\lambda_{H\chi \chi}^2$;
therefore, the ratio  $r_\chi= \Gamma (H \rightarrow \chi \chi)/\sigma^{\rm
SI}_{\chi p}$ depends only on the dark matter mass $M_\chi$ and known masses and
couplings (throughout, we assume the Higgs mass be $M_H\! = \! 125$ GeV).  This
allows us to relate the invisible Higgs branching fraction to the 
direct detection cross section:
\beq
{\rm BR}^{\rm inv}_\chi \equiv \frac{ \Gamma (H \rightarrow \chi \chi)}
{\Gamma^{\rm SM}_H + \Gamma (H \rightarrow \chi \chi)} = 
\frac{\sigma^{\rm SI}_{\chi p}}{\Gamma^{\rm SM}_H/r_\chi + \sigma^{\rm SI}_{\chi p}}
\eeq
with $\Gamma^{\rm SM}_H$ the total decay width into all particles in the SM. 
For a given $M_\chi$, the above formula connects the invisible branching
fraction probed at the LHC to the dark matter-nucleon scattering  cross section
probed by XENON100.   For  $m_p \ll M_\chi \ll \frac12 M_H$, and assuming the visible 
decay width equals to the SM total width $\Gamma^{\rm SM}_H=4.0$ MeV
\cite{hdecay}, one can write down the approximate relations in the three cases 
that we are considering,   
\bea
{\rm BR}^{\rm inv}_{S} &\simeq&\frac{ \left(\frac{\sigma^{\rm SI}_{S p}}{10^{-9} \mrm{pb}}\right)}{400 
\left(\frac{10\; \mrm{GeV}}{M_S}\right)^2 + \left(\frac{ \sigma^{\rm SI}_{S p}}{10^{-9}\mrm{pb}} \right)}
\nonumber
\\
{\rm BR}^{\rm inv}_{V}&\simeq&\frac{ \left(\frac{ \sigma^{\rm SI}_{V p}}{10^{-9} \mrm{pb}}\right)}{4 \times 10^{-2} 
\left(\frac{M_V}{10\; \mrm{GeV}}\right)^2 + \left( \frac{ \sigma^{\rm SI}_{V p}}{10^{-9}\mrm{pb}} \right)}
\nonumber
\\
{\rm BR}^{\rm inv}_{f}&\simeq&\frac{ \left(\frac{ \sigma^{\rm SI}_{f p}}{10^{-9} \mrm{pb}}\right)}{3.47
 + \left(\frac{ \sigma^{\rm SI}_{f p}}{10^{-9}\mrm{pb}} \right)}
 \label{Eq:brmax}
\eea
Thus, for a given mass of dark matter, an upper bound on the Higgs invisible
branching fraction  implies  an upper bound on the dark matter scattering cross
section on nucleons.    In Fig.~\ref{Fig:sigma40} we show the maximum allowed values of the
scattering cross section,  assuming the 20\% bound on  ${\rm BR}^{\rm inv}_\chi$, as follows from indirect constraints on the invisible width discussed in the previous section.   
Clearly, the relation between the invisible branching fraction and  the direct detection cross section
strongly depends  on  the spinorial nature of the dark matter particle, in particular, the strongest (weakest) bound is derived in the vectorial  (scalar)
case.  

In all cases,  the derived bounds on  $\sigma^{\rm SI}_{\chi p}$ are stronger
than the direct one from XENON100 in the entire range where $M_\chi \ll \frac12
M_H$.  In other words, the LHC is currently the most sensitive dark matter
detection apparatus, at least in the context of simple Higgs-portal models (even more so if $\chi$ is a pseudoscalar, as in \cite{LopezHonorez:2012kv}).  
This conclusion does not rely on the assumption that the present abundance of
$\chi$ is a thermal relic fulfilling the WMAP constraint of $\Omega_{DM} =
0.226$ \cite{WMAP}, and would only be stronger if $\chi$ constitutes only a
fraction of dark matter in the universe.
We also compared  the bounds to the
projected future sensitivity of the XENON100 experiment (corresponding to 60,000 kg-d, 5-30 keV and 45\% efficiency).  

Of course,  for $M_\chi > \frac12 M_H$, the Higgs boson  cannot decay into
dark matter\footnote{In this case, one should consider the pair production  of
dark matter particles through  virtual Higgs boson exchange, $pp\! \to\!  H^* X
\! \to \!\chi \chi X$. The rates are expected to be rather small 
\cite{Djouadi:2011aa}.}, in which case the LHC cannot  compete with
the XENON bounds.

\vspace*{-3mm}
\section*{Conclusions}
\vspace*{-3mm}

We have shown that monojet searches at the LHC already provide interesting
limits on invisible Higgs decays,   constraining the invisible rate to be less than the total SM Higgs production rate at the 95\% CL.   
This provides an important constrain on the models where the Higgs production cross section
is enhanced  and the invisible branching fraction is significant. 
Monojets searches are sensitive mostly to the gluon--gluon fusion  production
mode and, thus, they can also  probe invisible Higgs decays in models where the
Higgs coupling to the electroweak gauge bosons  is suppressed.  The limits could
be significantly improved when more data at higher center of mass energies are
collected,  provided systematic errors on the Standard Model contribution to the
monojet background can be reduced.  

We also analyzed in a model--independent way the interplay between the 
invisible Higgs branching fraction and the dark matter scattering cross section
on nucleons, in the context of  effective Higgs portal models.  The limit ${\rm
BR}_{\rm inv} < 0.2$, suggested by the combination of Higgs data in the visible
channels, implies a limit on the direct detection cross section that is stronger
than the current bounds from XENON100, for scalar, fermionic, and vectorial dark
matter alike. Hence, in the context of Higgs-portal models,  the LHC is
currently the most sensitive dark matter detection apparatus.

\noindent {\bf Acknowledgements:}   The authors would like  to thank J.~Baglio,
E.~Bagnaschi,  E.~Bragina, C.~Grojean, P. Boucaud,  and T.~Volansky as well as the Magic
Monday Journal Club for discussions. This  work was supported by the French ANR
TAPDMS {\it ANR-09-JCJC-0146}  and the Spanish MICINNÕs Consolider-Ingenio 2010
Programme  under grant  Multi- Dark {\it CSD2009-00064}. A.D. thanks the CERN TH
unit for hospitality and support.

\end{document}